\documentclass{ws-procs9x6}

\newcommand{\A}{{\mathfrak A}}
\newcommand{\Ao}{{\mathfrak A}_0}
\newcommand{\C}{{\cal C}}
\newcommand{\Nc}{{\cal N}}

\newcommand{\mc}{\mathcal}

\newcommand{\be}{\begin{equation}}
\newcommand{\en}{\end{equation}}
\newcommand{\D}{{\mc D}}
\newcommand{\restr}{\upharpoonright}

\newcommand{\LL}{\mc L}
\newcommand{\B}{{\cal B}}
\newcommand{\F}{{\cal F}}

\newcommand{\Lc}{{\cal L}}
\newcommand{\1}{1 \!\! 1}
\newcommand{\LD}{{\LL}^\dagger (\D)}
\newcommand{\G}{{\cal G}}

\newcommand{\N}{\mathbb N}
\newcommand{\Hil}{\mc H}

\newcommand{\bd}{\begin{displaymath}}

\newcommand{\h}{{\mathfrak H}}

\newcommand{\htil}{\widetilde{\mathfrak H}}

\newcommand{\lh}{{\mathcal L}({\mathfrak H})}

\newcommand{\hs}{\mathcal B_2 (\h )}
\newcommand{\LDl}{{\LL}_l^\dagger (\D)}
\newcommand{\LDr}{{\LL}_r^\dagger (\D)}

\newcommand{\bPhi}{\mbox{\boldmath $\Phi$}}
\newcommand{\bPsi}{\mbox{\boldmath $\Psi$}}

\newcommand{\bH}{\mathbf H}

\newcommand{\bP}{\mathbf P}

\newcommand{\bea}{\begin{eqnarray}}
\newcommand{\ena}{\end{eqnarray}}

\newcommand{\beano}{\begin{eqnarray*}}
\newcommand{\enano}{\end{eqnarray*}}

\begin{document}

\title{Modular Structures and Landau Levels}

\author{F. Bagarello$^*$}

\address{Dipartimento di Metodi e Modelli
Matematici, Universit\`a di Palermo,\\
Palermo, I-90128, Italy\\
$^*$E-mail: bagarell@unipa.it\\
www.unipa.it/$^\sim$bagarell}



\begin{abstract}

We review some recent results concerning Landau levels and Tomita-Takesaki modular theory. We also extend the general framework behind this to quasi *-algebras, to take into account the possible appearance of unbounded observables.

\end{abstract}


\bodymatter

\section{Introduction and mathematical theory}\label{sec:intro}
In a recent paper \cite{abh}, together with Ali and Honnouvo we have constructed an explicit application of the Tomita-Takesaki's modular theory to the so-called Landau levels. Here we review these results and we further extend the general settings to consider the presence also of unbounded operators.

The motion of a quantum electron in a constant electromagnetic field produces energy (Landau) levels which are linearly
spaced and infinitely degenerate. The
Hamiltonian of the system, using "clever" variables, can be written as the hamiltonian of a single harmonic oscillator. Moreover, if the sense of the magnetic field is reversed,
one obtains a second Hamiltonian, similar to the first one, but commuting with it. Both these
Hamiltonians can be written in terms of two pairs of mutually commuting oscillator type
creation and annihilation operators, which then
generate two von Neumann algebras which mutually commute, and in fact are commutants
of each other. This leads to the existence of a modular
structure, in the sense of the Tomita-Takesaki theory \cite{takesaki,aliemch,emch} .

\subsection{Summary of the  mathematical theory}\label{sec:summ-math}

We will now quickly review some basic facts of the Tomita-Takesaki modular theory of
von Neumann algebras.
Details and proofs of statements may
be found, for example,  in \cite{sewell,stratila,takesaki} .  Let $\mathfrak A$ be a von
Neumann algebra on a Hilbert space $\h$ and $\mathfrak A^\prime$ its commutant. Let
$\bPhi \in \h$ be a unit vector which is cyclic and separating for $\mathfrak A$.  Then
the corresponding state $\varphi$ on the algebra, $\langle \varphi\; ; \; A\rangle =
\langle \bPhi \mid A \bPhi \rangle\; , \; A \in \mathfrak A\; , $ is faithful and normal.
Consider the antilinear map,
\be
  S:\h \longmapsto \h\; , \qquad SA\bPhi = A^*\bPhi\; , \; \forall A \in \mathfrak A \; .
\label{mod-antilin-map}
\en
Since $\bPhi$ is cyclic, this map is densely defined and in fact it can be shown that it is
closable. We denote its closure again by $S$ and  write its polar decomposition as
\be
  S = J\Delta^\frac 12 = \Delta^{-\frac 12}J\; , \quad \text{with} \quad \Delta = S^*S\; .
\label{mod-pol-decomp}
\en
The operator $\Delta$, called the {\em modular operator}, is positive and self-adjoint. The
operator $J$, called the {\em modular conjugation operator}, is antiunitary and satisfies
$J = J^*\; , \; J^2 = I_\h$. Note that the antiunitarity of $J$ implies that
$\langle J\phi \mid J\psi \rangle = \langle \psi \mid \phi\rangle\; , \; \forall \phi, \psi
\in \h$.

  Since $\Delta$ is self-adjoint, using its spectral representation we see that for $t\in
\mathbb R$ the family of operators $\Delta^{-\frac {it}\beta}$, for some fixed $\beta > 0$,
defines a unitary family of automorphisms of the algebra $\mathfrak A$. Denoting these
automorphisms by $\alpha_\varphi (t)$, we may write
\be
  \alpha_\varphi (t)[A] = \Delta^{\frac {it}\beta}A\Delta^{-\frac {it}\beta}\; , \;\;
  \forall A \in \mathfrak A\; .
\label{mod-automorph-grp}
\en
Thus, they constitute a strongly continuous one-parameter group of automorphisms, called the
{\em modular automorphism group\/}.
Denoting the generator of this one-parameter group by $\mathbf H_\varphi$, we
get
\be
   \Delta^{-\frac {it}\beta} = e^{it\mathbf H_\varphi} \quad \text{and} \quad
     \Delta = e^{-\beta \mathbf H_\varphi}\; .
\label{mod-automorph-grp2}
\en
It can then be shown that the state $\varphi$ is invariant under this automorphism group, that
\be
  e^{-\beta \mathbf H_\varphi}\bPhi = \bPhi\; ,
\qquad  \Delta^{\frac {it}\beta}\;\mathfrak A \;
    \Delta^{-\frac {it}\beta} = \mathfrak A\; ,
\en
and that the antilinear map $J$ interchanges $\mathfrak A$ with its commutant $\mathfrak A^\prime$:
\be
  J\mathfrak A J = \mathfrak A^\prime\; .
\label{mod-interch}
\en

  Finally, the state $\varphi$ can be shown to satisfy the {\em KMS (Kubo-Martin-Schwinger)
condition\/}, with respect to the automorphism group $\alpha_\varphi (t)\; , \; t \in
\mathbf R$, in the following sense. For any two $A,B \in \mathfrak A$, the  function
\be
   F_{A, B} (t) = \langle\varphi\; ;\; A\; \alpha_\varphi (t)
   [B]\rangle\; ,
\label{mod-KMS-func}
\en
has an extension to the strip $\{ z = t+ iy\mid t \in \mathbb R,\; y \in [0, \beta]\} \subset
\mathbb C$ such that $F_{A, B} (z)$  is analytic in the open strip $(0, \beta )$ and
continuous on its boundaries. Moreover, it also satisfies the boundary condition ({\em at inverse
temperature} $\beta$),
\be
  \langle\varphi\; ;\; A\; \alpha_\varphi (t + i\beta ) [B]\rangle =
   \langle\varphi\; ;\; \alpha_\varphi (t) [B]  \;A\rangle \; , \quad t \in \mathbb R\; .
\label{mod-KMS-prop}
\en

\subsubsection{The role of the  Hilbert-Schmidt operators}\label{sec:setup}

A simple example of the Tomita-Takesaki theory and its
related KMS states can be built on the space of Hilbert-Schmidt operators on a Hilbert space.
The set of Hilbert-Schmidt operators is by itself a Hilbert space, and there are two
preferred algebras of operators on it, which carry the modular structure. The presentation
here follows that in \cite{aag_book} (Chapter 8, Section 4).

Again,  let $\h$ be a (complex, separable) Hilbert space of dimension $N$  (finite or infinite) and
$\{\zeta_i\}_{i=1}^N$  an orthonormal basis of it ($\langle \zeta_i \mid \zeta_j \rangle =
\delta_{ij}$). We denote by $\mathcal B_2 (\h )$ the
space of all Hilbert-Schmidt operators on $\h$. This is a Hilbert space with scalar
product
$$ \langle X\mid Y\rangle_2 = \text{Tr}[X^* Y]\; . $$
The vectors
\be \{X_{ij} = \vert\zeta_i\rangle\langle \zeta_j \vert \mid i,j = 1,2,\ldots , N\}\; ,
\label{basis-vects}
\en
form an orthonormal basis of  $\mathcal B_2 (\h ) $:
$$ \langle X_{ij}\mid X_{k\ell}\rangle_2  = \delta_{ik}\delta_{\ell j}\; . $$
In particular, the vectors
\be
    \mathbb P_i = X_{ii} = \vert\zeta_i \rangle\langle \zeta_i \vert \; ,
\label{proj-op1}
\en
are one-dimensional projection operators on $\h$. In what follows
 $I$ will denote the identity operator on $\h$ and  $I_2$ that on $\hs$.

   We identify a special class of linear operators on $\mathcal B_2 (\h ) $,
denoted by $A\vee B , \; A,B \in \mathcal L (\h)$, which act on a vector $X \in
\mathcal B_2 (\h ) $ in the manner:
$$    (A\vee B) (X) = AXB^*\; . $$
Using the scalar product in $\hs$, we see that
$$ \rm{Tr} [X^* (AYB^*) ] = \rm{Tr} [(A^*X B)^*  Y) ] \Longrightarrow
  (A\vee B)^* = A^* \vee B^* \; , $$
  and since for any $X \in \hs$
  $$ (A_1\vee B_1) (A_2 \vee B_2) (X) = A_1 [(A_2 \vee B_2 )(X)]B_1^*
    = A_1 A_2 X B_2^* B_1^* \; , $$
we have,
  \be
  (A_1\vee B_1) (A_2 \vee B_2) = (A_1A_2 ) \vee (B_1B_2) \; .
\label{comp-law}
\en

  There are two special von Neumann algebras which can be built out of these operators.
These are
\be
   \mathfrak A_\ell = \{ A_\ell = A \vee I \mid A \in \mathcal L (\h)\} \; , \qquad
   \mathfrak A_{\rm{r}} = \{ A_{\rm{r}} = I \vee A \mid A \in \mathcal L (\h)\} \; .
\label{left-right-alg}
\en
They are mutual commutants and both are factors:
\be
  (\mathfrak A_\ell)^\prime =  \mathfrak A_{\rm{r}} \; , \qquad
  ( \mathfrak A_{\rm{r}})^\prime  =  \mathfrak A_\ell\; , \qquad
  \mathfrak A_\ell \cap \mathfrak A_{\rm{r}} = \mathbb C I_2\; .
\label{factors}
\en

Consider now the operator $J: \hs \longrightarrow \hs$, whose action on the vectors
$X_{ij}$ in (\ref{basis-vects}) is given by
\be
   JX_{ij} = X_{ji} \Longrightarrow J^2 = I_2 \quad \text{and} \quad J(\vert \phi
   \rangle\langle \psi \vert )  = \vert\psi \rangle\langle \phi\vert\;,
   \quad \forall \phi , \psi \in \h\; .
\label{J-op}
\en
This operator is antiunitary, and since
$$ [J(A\vee I) J]X_{ij} = J (A\vee I)X_{ji} = J(AX_{ji}) = J(A\vert\zeta_j\rangle\langle
\zeta_i\vert) =$$
$$= \vert \zeta_i\rangle \langle \zeta_j\vert A^*  = (I \vee A )X_{ij} \; , $$
we immediately get
\be
  J \mathfrak A_\ell J  =  \mathfrak A_{\rm{r}}\; .
\label{J-op2}
\en

A KMS state can be introduced starting from a set of non-zero, positive
numbers $\alpha_i \; , \;\; i=1,2, \ldots , N$,
 satisfying
$\sum_{i=1}^N \alpha_i = 1$. Indeed, let
\be
  \bPhi =  \sum_{i=1}^N \alpha_i^{\frac 12}\; \mathbb P_i =
   \sum_{i=1}^N \alpha_i^{\frac 12}\; X_{ii} \in  \hs\; .
\label{KMS-vect}
\en
Then $\bPhi$ defines a vector state $\varphi$ on the von Neumann algebra
$\mathfrak A_\ell$: for any $A\vee I \in \mathfrak A_\ell$,
we put
\be
 \langle \varphi \; ; \; A\vee I \rangle  =
 \langle \bPhi \mid  (A\vee I)( \bPhi) \rangle_2  =
     \text{Tr} [\bPhi^* A \bPhi ] = \text{Tr}[\rho_\varphi A ],
\label{vect-state}
\en
with $\rho_\varphi = \sum_{i=1}^N\alpha_i \mathbb P_i$. Moreover $\bPhi$ is cyclic and separating for $\mathfrak A_\ell$, while  $\varphi$ is faithful and normal \cite{abh} .

It is also possible to show that the state $\varphi$ is indeed a KMS state for a particular choice of $\alpha_i$.
For that, we first need to construct a time evolution $\alpha_\varphi (t), \; t \in \mathbb R$, on the algebra
$\mathfrak A_\ell$, using the state $\varphi$, with respect to which it has the KMS property,
for fixed $\beta >0$,
\be
  \langle\varphi\; ;\; A_\ell\; \alpha_\varphi (t + i\beta ) [B_\ell]\rangle =
   \langle\varphi\; ;\; \alpha_\varphi (t) [B_\ell]  \;A_\ell\rangle \; , \quad \forall
   A_\ell, B_\ell \in  \mathfrak A_\ell\; ,
\label{KMS-prop}
\en
and moreover the function,
\be
   F_{A_\ell, B_\ell} (z) = \langle\varphi\; ;\; A_\ell\; \alpha_\varphi (z)
   [B_\ell]\rangle\; ,
\label{KMS-func}
\en
is analytic in the strip $\{\Im (z) \in (0, \beta)\}$ and continuous on its boundaries. We
start by defining the operators,
\be
  \bP_{ij} = \mathbb P_i \vee \mathbb P_j \; , \qquad i,j = 1,2, \ldots , N\;
\label{proj op2}
\en
where the $\mathbb P_i$ are the projection operators on $\h$ defined in (\ref{proj-op1}).
Clearly,
the $\bP_{ij}$ are projection operators on the Hilbert space $\hs$.

   Using $\rho _\varphi$ in (\ref{vect-state}) and for a fixed $\beta > 0$, we define the
   operator
 $H_\varphi$ as:
 \be
  \rho_\varphi = e^{-\beta H_\varphi} \Longrightarrow H_\varphi =
  -\frac 1\beta \sum_{i=1}^N \left(\ln \alpha_i\right) \mathbb P_i\; .
 \label{hamiltonian1}
 \en
 Next we define the operators:
 \be
  H_\varphi^\ell = H_\varphi \vee I\; , \qquad  H_\varphi^{\rm{r}} =
  I\vee H_\varphi \; , \qquad
  \bH_\varphi =  H_\varphi^\ell - H_\varphi^{\rm{r}} \; ,
 \label{hamiltonian2}
 \en
 Since $\sum_{i=1}^N \mathbb P_i = I$, we may also write
$$
  H_\varphi^\ell = -\frac 1\beta \sum_{i,j =1}^N \ln \alpha_i \bP_{ij}\; , \quad
\text{and} \quad
H_\varphi^{\rm{r}} = -\frac 1\beta \sum_{i,j =1}^N \ln \alpha_j \bP_{ij}\; . $$
Thus,
\be
 \bH_\varphi =  -\frac 1\beta \sum_{i,j =1}^N \ln \left[\frac {\alpha_i}{\alpha_j}\right ]
    \bP_{ij}\; .
\label{hamilton3}
\en
Using the operator.
\be
   \Delta_\varphi : = \sum_{i,j =1}^N \left[\frac {\alpha_i}{\alpha_j}\right ] \bP_{ij}
= e^{-\beta \bH_\varphi}\; ,
\label{delta-op}
\en
we further define a time evolution operator on $\hs$:
\be
 e^{i\bH_\varphi t } = [\Delta_\varphi ]^{-\frac {it}\beta}\; . \qquad t \in \mathbb R\; ,
\label{timeev-op}
\en
and we note that, for any $X\in \hs$,
\beano
 e^{i\bH_\varphi t } (X) & = & \sum_{i,j =1}^N
\left[\frac {\alpha_i}{\alpha_j}\right ]^{-\frac {it}\beta} \bP_{ij}(X) =
\left[\sum_{i=1}^N(\alpha_i)^{-\frac {it}\beta}\mathbb P_i\right]\vee
 \left[\sum_{j=1}^N(\alpha_j)^{-\frac {it}\beta}\mathbb P_j (X)\right]\\
 & = &  e^{iH_\varphi t } (X) e^{-iH_\varphi t }\; ,
\enano
so that
\be
  e^{i\bH_\varphi t } =  e^{iH_\varphi t } \vee e^{iH_\varphi t }\; ,
\label{timeev-op2}
\en
where $H_\varphi$ is the operator introduced in (\ref{hamiltonian1}). From the
definition of the vector $\bPhi$ in (\ref{KMS-vect}), it is clear that it commutes with
$H_\varphi$ and hence that it is invariant  under this time evolution:
\be
 e^{i\bH_\varphi t } (\bPhi)  = e^{iH_\varphi t }\; \bPhi \; e^{-iH_\varphi t }
  = \bPhi \; .
\label{inv-vect}
\en

  Finally, using $e^{i\bH_\varphi t }$ we define the time evolution $\alpha_\varphi$
on the algebra  $\mathfrak A_\ell$, in  the manner (see (\ref{mod-automorph-grp})):
\be
  \alpha_\varphi (t) [A_\ell ] = e^{i\bH_\varphi t }\; A_\ell \; e^{-i\bH_\varphi t } \qquad
 \forall A_\ell  \in \mathfrak A_\ell \; .
\label{timeev-op3}
\en
Writing $A_\ell = A\vee I\; , \;\; A \in \lh$, and  using the composition law (\ref{comp-law}),
we see that
\be
 e^{i\bH_\varphi t }\; A_\ell \; e^{-i\bH_\varphi t } =
\left[e^{iH_\varphi t }\; A \; e^{-iH_\varphi t }\right]\vee I\; ,
\label{timeev-op4}
\en
so that, by virtue of (\ref{vect-state}),
\be
 \langle \varphi\; ; \;  \alpha_\varphi (t) [A_\ell ]\rangle =  \text{Tr}\left[ \rho_\varphi \;
   e^{iH_\varphi t }\; A \; e^{-iH_\varphi t }\right] =
   \langle \varphi\; ; \;  A_\ell \rangle\; ,
\label{inv-state}
\en
since $\rho_\varphi$ and $H_\varphi$ commute. Thus, the state $\varphi$ is invariant
under the time evolution $\alpha_\varphi$. We finally refer to \cite{abh} for the proof of the KMS condition.

  We now analyze the antilinear operator $S_\varphi : \hs \longrightarrow \hs$, which acts as
(see (\ref{mod-antilin-map}))
\be
  S_\varphi (A_\ell \bPhi ) = A_\ell^* \bPhi \; , \qquad \forall A_\ell \in \mathfrak A_\ell\; .
\label{antilinop}
\en
Taking $A_\ell = A\vee I$,
$$
S_\varphi (A_\ell \bPhi ) = A_\ell^* \bPhi \; , \quad \forall A_\ell \in \mathfrak A_\ell
\quad \Longleftrightarrow \quad
S_\varphi (A\bPhi ) = A^* \bPhi \; , \quad \forall A \in \lh\; . $$
Using (\ref{KMS-vect}) we may write,
$$ S_\varphi (A\bPhi ) = A^* \bPhi \quad \Longrightarrow \quad
\sum_{i=1}^N \alpha_i^{\frac 12}S_\varphi (A\mathbb P_i ) =
\sum_{i=1}^N \alpha_i^{\frac 12} A^* \mathbb P_i \; . $$
Taking  $A = X_{k \ell}$ (see (\ref{basis-vects})) and using $X_{k\ell}\mathbb P_i
= \delta_{\ell i }X_{ki}$, we then get
\be
  \alpha_\ell^{\frac 12}S_\varphi (X_{k\ell} ) = \alpha_k^{\frac 12}S_\varphi (X_{\ell  k} )
  \quad \Longrightarrow \quad  S_\varphi (X_{k\ell} ) =
  \left[\frac { \alpha_k}{ \alpha_\ell}\right]^{\frac 12}X_{\ell k}\; .
\label{antiliop2}
\en
Since any $A \in \lh$ can be written as $A = \sum_{i,j =1}^N a_{ij}X_{ij}$, where
$a_{ij} = \langle \zeta_i \mid A\zeta_j \rangle$, and furthermore, since
$ \bP_{ij} (X_{k\ell}) = X_{ij}\delta_{ik}\delta_{j\ell}$, we obtain using
(\ref{J-op}) and  (\ref{delta-op}),
\be
   S_\varphi = J [\Delta_\varphi]^{\frac 12}\; ,
\label{antilinop3}
\en
which in fact, also gives the polar decomposition of $S_\varphi$.

\section{Application to Landau levels}\label{sec:applications}
  We now show how the above setup, based on $\hs$, can be applied to a specific physical
situation namely, to the case of  an electron subject to a constant magnetic
field, as discussed in \cite{alibag} .

In that case, $\h = L^2 (\mathbb R)$ and
the mapping $\mathcal W : \hs  \longrightarrow L^2 (\mathbb R^2, dx\;dy)$, with
\be
   (\mathcal W X)(x,y) = \frac 1{(2\pi)^{\frac 12}} \text{Tr}[ U(x, y)^* X], \quad
\text{where} \quad U(x,y) = e^{-i(xQ + yP)},
\label{wigmap}
\en
$Q, P$ being the usual position and momentum operators ($[Q,P] = iI$), transfers the whole
modular structure unitarily to the Hilbert space $\htil = L^2 (\mathbb R^2, dx\;dy)$. The
mapping $\mathcal W$ is often referred to as the {\em Wigner transform} in the physical
literature.

To work this out in some detail, we start by  constructing the
Hamiltonian $H_\varphi$ (see (\ref{hamiltonian1})),  using the oscillator
Hamiltonian $H_\text{osc} = \frac 12 (P^2 + Q^2)$ on $\h$.  Let us  choose the
orthonormal basis set of vectors $\zeta_n\; , n =0,1, 2, \ldots \infty$, to be the
eigenvectors of $H_\text{osc}$:
\be
  H_\text{osc} \zeta_n = \left(n + \frac 12 \right) \zeta_n\; .
\label{osc-eigenfunc}
\en
As it is well known, the $\zeta_n$ are the Hermite functions,
\be
  \zeta_n (x) = \frac 1{\pi^{\frac 14}}\frac 1{\sqrt{2^n \;n!}} \;e^{-\frac {x^2}2} h_n (x)\; ,
\label{herm-func}
\en
 the $h_n$ being the Hermite polynomials, obtainable as:
\be
  h_n (x) = (-1)^n \; e^{x^2} \partial^n_x \; e^{- x^2}\; .
\label{real-herm-poly}
\en

Consider now the operator $e^{-\beta H_\text{osc}}$, for some fixed $\beta > 0$. We have
$$
   e^{-\beta H_\text{osc}} = \sum_{n=0}^\infty e^{-(n + \frac 12)\beta}\mathbb P_n
 \quad \text{and} \quad  \text{Tr} \left[ e^{-\beta H_\text{osc}} \right ] =
 \frac {e^{-\frac {\beta}2}}{1 - e^{-\beta}}\; .  $$
 Thus we take
\be
  \rho_\varphi = \frac {e^{-\beta H_\text{osc}}}{\text{Tr}
  \left[ e^{-\beta H_\varphi} \right ]} =
  (1 - e^{-\beta})\sum_{n=0}^\infty e^{-n\beta}\mathbb P_n,\; \, \text{and}
  \, \bPhi = \left[1 - e^{-\beta}\right]^{\frac 12} \sum_{n=0}^\infty
  e^{-{\frac n2}\beta}\mathbb P_n\; .
\label{KMS4}
\en
Following  (\ref{vect-state}) and (\ref{hamiltonian1}) we write
$$
    \rho_\varphi = \sum_{n=0}^\infty \alpha_n \mathbb P_n \; , \qquad
    \alpha_n =  (1 - e^{-\beta})e^{-n\beta}\; , $$
and
\bea
  H_\varphi & =&  - \frac 1\beta \sum_{n=0}^\infty \ln\left[(1 - e^{-\beta})e^{-n\beta}\right]
      \mathbb P_n
    =    \sum_{n=0}^\infty \left[ n - \frac { \ln(1 - e^{-\beta})}\beta \right]
      \mathbb P_n \nonumber\\
     & = &  H_\text{osc} - \left[\frac 12 +  \frac {\ln(1 - e^{-\beta})}\beta \right]  I\; ,
\label{modhamilt}
  \ena
which is the Hamiltonian giving the time evolution  $\alpha_\varphi (t)$,  with respect
to which the above $\rho_\varphi$ defines the KMS state $\varphi$. Since the difference
between  $H_\varphi$ and $H_{\text{osc}}$ is just a constant,
we shall identify these two Hamiltonians in the sequel.

  As stated earlier, the dynamical model that we consider is that of a single electron
of unit charge, placed in the $xy$-plane and subjected to a constant magnetic
field, pointing along the {\em positive $z$-direction\/}. The classical
Hamiltonian of the system, in  convenient units, is
\be
  H_\text{elec} = \frac 12 (\vec p - \vec A )^2 = \frac 12 \left(p_x + \frac y2 \right)^2 +
      \frac 12 \left(p_y - \frac x2 \right)^2,
\label{elec-ham}
\end{equation}
where we have chosen the magnetic vector potential to be $\vec A^\uparrow := \vec A =
\frac 12 (-y, x, 0)$ (so that the magnetic field is $\vec B = \nabla \times \vec A^\uparrow
= (0,0, 1)$).

Next, on  $\htil = L^2 (\mathbb R^2 , dxdy )$, we introduce the quantized
observables, \be
 p_x + \frac y2 \longrightarrow Q_- = -i\frac \partial{\partial x } + \frac y2 \; , \qquad
 p_y - \frac x2 \longrightarrow P_- = -i\frac \partial{\partial y } - \frac x2 \; ,
\label{quant-obs1}
\end{equation}
which satisfy $[Q_-, P_- ] = iI_{\htil}$ and in terms of which the quantum Hamiltonian
corresponding to $H_\text{elec}$ becomes
\be
  H^\uparrow = \frac 12 \left( P_-^2 + Q_-^2\right)\; .
\label{qu-elec-ham1}
\end{equation}
This is the same as the oscillator Hamiltonian in one dimension, $H_{\text{osc}}$,
given above (and hence the same as $H_\varphi$ in (\ref{modhamilt}), with our convention
of identifying these two). The eigenvalues of this Hamiltonian are then
$E_\ell = (\ell + \frac 12 ), \;
\ell =0,1,2, \ldots \infty$. However, this time each level is infinitely degenerate,
and we will denote the corresponding normalized eigenvectors by
$\Psi_{n \ell}$, with $\ell = 0, 1,2, \ldots , \infty$, indexing  the energy level and
$n = 0,1,2, \ldots , \infty$, the degeneracy at each level. If the magnetic field
were aligned along the {\em negative
$z$-axis} (with $\vec A^\downarrow = \frac 12 (y, -x, 0)$ and
$\vec B = \nabla \times \vec A^\downarrow = (0,0, -1)$), the corresponding quantum
Hamiltonian would have been
\be
  H^\downarrow = \frac 12 \left( P_+^2 + Q_+^2\right)\; .
\label{qu-elec-ham2}
\end{equation}
with
\be
 Q_+ = -i\frac \partial{\partial y } + \frac x2 \; , \qquad
 P_+ = -i\frac \partial{\partial x } - \frac y2 \; ,
\label{quant-obs2}
\end{equation}
and $[Q_+, P_+ ] = iI_{\htil}$. The two sets of operators $\{ Q_\pm , P_\pm\},$
mutually commute:
\be
  [Q_+ , Q_- ] = [P_+ , Q_- ] = [Q_+ , P_- ] = [P_+ , P_- ] = 0 \; .
\label{commutants}
\end{equation}

Thus, $[H^\downarrow , H^\uparrow ] = 0$ and the eigenvectors $\Psi_{n \ell}$
of $H^\uparrow$ can be  chosen so that they are
also the eigenvectors of $H^\downarrow$ in the manner
\be
  H^\downarrow\Psi_{n\ell} =  \left(n + \frac 12 \right)\Psi_{n\ell} \; , \qquad
  H^\uparrow\Psi_{n\ell} =  \left(\ell + \frac 12 \right)\Psi_{n\ell}\; ,
\label{deg-lift}
\end{equation}
so that $H^\downarrow$ lifts the degeneracy of $H^\uparrow$ and vice versa.


Then, it is well known (see, for example, \cite{aag_book}) that the  map $\mathcal W$ in
(\ref{wigmap})
is unitary and straightforward computations (see, for example \cite{alibag}) yield,
\be
 \mathcal W \begin{pmatrix} Q\vee I_\h \\ P\vee I_\h \end{pmatrix}\mathcal W ^{-1} =
 \begin{pmatrix} Q_+ \\ P_+ \end{pmatrix}\; , \qquad
 \mathcal W \begin{pmatrix} I_\h \vee Q \\ I_\h \vee P \end{pmatrix} \mathcal W^{-1} =
 \begin{pmatrix} Q_- \\ P_- \end{pmatrix}\;,
\label{alg-transf1}
\end{equation}
and
\be
\mathcal W \begin{pmatrix} H_\text{osc}\vee I_\h \\ I_\h\vee H_\text{osc}
\end{pmatrix}\mathcal W^{-1} =
 \begin{pmatrix} H^\downarrow \\ H^\uparrow \end{pmatrix}\; , \qquad \mathcal W  X_{n\ell} = \Psi_{n\ell },
\label{alg-transf2}
\end{equation}
where the $X_{n\ell}$ are the basis vectors defined in
(\ref{basis-vects}) and the $\Psi_{n\ell}$ are the normalized
eigenvectors defined in (\ref{deg-lift}). This also means that
these latter vectors form a basis of $\htil = L^2(\mathbb R^2 , dxdy)$. Finally, note that the
two sets of operators, $\{Q_+ , P_+  \}$ and $\{Q_- , P_- \}$, generate
the two von Neumann algebras
$\mathfrak A_+$ and $\mathfrak A_-$, respectively, with $\mathcal W \mathfrak A_\ell
\mathcal W^{-1} =
\mathfrak A_+$ and $\mathcal W \mathfrak A_{\text{r}}\mathcal W^{-1}  = \mathfrak A_-$.
Thus physically, the two commuting algebras correspond to the two directions of the
magnetic field. The KMS state $\bPsi = \mathcal W \bPhi$, with $\bPhi$ given by
(\ref{KMS4}) is just the {\em Gibbs equilibrium state} for this physical system.

\subsection{A second representation}\label{sec:second-rep}
  It is interesting to pursue this example a bit further using different variables in the description of the system.
As before, let us consider the electron in a uniform magnetic field
oriented in the positive $z$-direction, with vector potential
$\vec A^\uparrow =\frac{1}{2}(-y,x,0)$ and magnetic field $\vec B = \nabla \times \vec A^\uparrow
= (0,0, 1)$). The classical Hamiltonian is  given by
$H^\uparrow=\frac{1}{2}\left(\vec p- \vec A^\uparrow\right)^2$.
There are several possible ways to write this Hamiltonian, which are more convenient
than using the coordinates $x, y$ and $z$. One such representation was used above and we indicate below a second possibility. Note that the
quantized Hamiltonian may be splitted into a free part $H_0$ and an interaction
or angular momentum part,
$H_{\text{int}}^\uparrow$:
\be
\left\{
\begin{array}{ll}
H^\uparrow=H_0+H_{\text{int}}^\uparrow,\\[2pt]
H_0=H_{0,x}+H_{0,y}=\frac{1}{2}\left( \widehat{p}_x^2+ \dfrac {\widehat{x}^2}4\right)+
\frac{1}{2}\left(\widehat{p}_y^2+ \dfrac {\widehat{y}^2}4 \right),\\[2pt]
H_{\text{int}}^\uparrow=- \dfrac 12 (\widehat{x}\widehat{p}_y- \widehat{y}\widehat{p}_x)
    =-\widehat{l}_z\; .
\end{array}
\right.\label{11}\en
with the usual definitions of $\widehat{x} , \widehat{p}_x\; ,$ etc. Of
course, $[\widehat{x}, \widehat{p}_x]=[\widehat{y} ,\widehat{p}_y]=iI_{\htil}$, while all
the other commutators are zero. Introducing the corresponding annihilation operators
\be
a_x= \frac 1{\sqrt{2}} [\widehat{x} + i\widehat{p}_x ]\;,\qquad
a_y= \frac 1{\sqrt{2}} [\widehat{y} + i\widehat{p}_y ]\; ,
\label{12}
\en
and their adjoints
\be
a_x^* = \frac 1{\sqrt{2}} [\widehat{x} -  i\widehat{p}_x]\; ,\qquad
a_y^* = \frac 1{\sqrt{2}} [\widehat{y} - i\widehat{p}_y] \; ,
\label{122}
\en
which satisfy the canonical commutation rules
$[a_x,a_x^*]=[a_y,a_y^*]=I_{\htil}$, while all the other commutators are zero,
the hamiltonian $H^\uparrow$ can be written as
$H^\uparrow=H_0+H_{\text{int}}^\uparrow$, with $H_0= \left(a_x^*a_x+a_y^*a_y+ I_{\htil}\right)$,
$H_{\text{int}}^\uparrow=-i(a_x a_y^*-a_ ya_x^*)$. $H^\uparrow$ does not appear to be
diagonal even in this form, so that another {\em change of variables} is required.

  Using the operators $Q_\pm , P_\pm$, given in (\ref{quant-obs1}) and (\ref{quant-obs2}),
let us define
\bea
  A_+  & =  & \frac 1{\sqrt{2}} (Q_+ + iP_+)  = \frac 34 (a_x - ia_y) - \frac 14 (a_x^* + ia_y^* )\; ,  \nonumber\\
  A_+^*  & = & \frac 1{\sqrt{2}} (Q_+ - iP_+)  = \frac 34 (a_x^* + ia_y^*) - \frac 14 (a_x - ia_y )\; ,
  \nonumber\\
 A_- & = &  \frac 1{\sqrt{2}} (iQ_- - P_-)  = \frac 34 (a_x + ia_y) - \frac 14 (a_x^* + ia_y^* )\; , \nonumber\\
   A_-^* &  = & \frac 1{\sqrt{2}} (-iQ_- - P_-)  = \frac 34 (a_x^* - ia_y^*) - \frac 14 (a_x - ia_y ) \; .
\label{new-ccr}
\ena

These satisfy the commutation relations,
\be
 [A_\pm , A^*_\pm ] = 1\; ,
\label{new-ccr2}
\en
with all other commutators being zero. In terms of these, we may write the
two Hamiltonians as (see (\ref{qu-elec-ham1}) and (\ref{qu-elec-ham2}),
\be
  H^\uparrow =  N_- + \frac 12 I_{\htil} \; , \quad
  H^\downarrow = N_+ + \frac 12 I_{\htil}\;, \quad \text{with} \quad N_\pm =
  A^*_\pm A_\pm\; .
\label{qu-elec-ham6}
\en
Furthermore,
\be
  H_0 = \frac 12 (N_+ + N_- +1) \quad \text{and} \quad H_{\text{int}}^\uparrow
    =  -\frac 12 (N_+ - N_-)\; , \quad H_{\text{int}}^\downarrow
    =  \frac 12 (N_+ - N_-)\; .
\label{qu-elec-ham7}
\en

   The eigenstates of $H^\uparrow$
are now easily written down. Let $\Psi_{00}$ be such that $A_-\Psi_{00}=A_+\Psi_{00}=0$.
Then we define
\be
\Psi_{n \ell}:=\frac{1}{\sqrt{n!\ell!}}\left(A_+^*\right)^n\left(A_-^*\right)^\ell\Psi_{0 0},
\label{15}
\en
where $n,\ell =0,1,2,\ldots$. All the relevant operators are now diagonal in this basis:
$N_+\Psi_{n \ell }=n\Psi_{n \ell }\;,\;\; N_-\Psi_{n \ell }=\ell \Psi_{n \ell }\;,\;\;
H_0\Psi_{n \ell }= \frac 12
(n+\ell+1)\Psi_{n \ell}$ and $H_{\text{int}}^\uparrow\Psi_{n \ell }=
\frac 12 (n-\ell )\Psi_{n \ell}$. Hence
\be
H^\uparrow\Psi_{n \ell }= \left(\ell +\frac{1}{2}\right)\Psi_{n \ell}.
\label{16}
\en
This means that, as already stated, each level $\ell$ is infinitely degenerate, with $n$ being the  degeneracy
index. Again, this degeneracy can be lifted in a
physically interesting way namely, by considering the {\em reflected} magnetic field with
vector potential $\vec A^\downarrow=\frac{1}{2}(y,-x,0)$,
with the magnetic field  directed along the negative $z$-direction.  The same electron
considered above is now described by the other  Hamiltonian,
$H^\downarrow$, which can be written as
\be
\left\{
\begin{array}{ll}
H^\downarrow=\frac{1}{2}\left(\vec p - \vec A^\downarrow\right)^2=
H_0+H_{\text{int}}^\downarrow,\\
H_{\text{int}}^\downarrow=-H_{\text{int}}^\uparrow
\end{array}
\right.\label{17}
\en
Thus, since  $H^\downarrow$ can also be written as  in (\ref{qu-elec-ham6}), its
eigenstates are again  the same $\Psi_{n \ell }$ given in (\ref{15}). (Recall that
 $[H^\uparrow,H^\downarrow]=0$, so that they can be simultaneously
diagonalized.) This also means that, as in (\ref{alg-transf2}), $\mathcal W X_{n\ell}
=\Psi_{n\ell}$ and the closure of the linear span of the $\Psi_{n \ell}$'s is the Hilbert space
$\htil = L^2 (\mathbb R^2 , dx\;dy )$.

It should be mentioned that in \cite{abh} yet another representation has been considered, relating the two different directions of the magnetic field to holomorphic and anti-holomorphic functions. We refer to \cite{abh} for more details on this subject, and in particular to the interesting possibility of introducing a map $\mathcal J$ which is, at the same time, (i) the modular map of the Tomita-Takesaki theory;
(ii) the complex conjugation map;
(iii) the map which reverses the uniform magnetic field,
  from $\vec B$ to $-\vec B$,
thus transforming  $\mathcal H^\uparrow$ to $\mathcal H^\downarrow$;
(iv) the operator  interchanging the two mutually commuting von Neumann
algebras  $\mathfrak U_\pm$;
(v) an interwining operator in the sense of \cite{intop}. Hence the structure is quite rich both from the mathematical and from the physical sides.

We also refer to \cite{abh} for the analysis of the coherent and bi-coherent states associated to the structure considered here.

\section{What for unbounded operators?}

It is well known that in many physical problems, mainly involving quantum systems with infinite degrees of freedom, unbounded operators play a crucial role \cite{bagrev}. This makes the above construction impossible to be used, as it is. The reason is that, in this case, algebras of operators cannot be introduced easily since unbounded operators cannot be multiplied without {\em danger}. Several partial structures have been introduced along the years, see \cite{bagrev} and references therein, which are essentially set of operators for which the multiplication is only partially defined. So, in order to extend what we have described above to this case, we have first to replace the Hilbert-Schmidt operators
with a different set with similar properties and which, moreover, contains unbounded operators. This can be  done with a little effort \cite{aitbook} . We begin with some notation:

  let $\A$ be a linear space,
$\Ao\subset\A$ a $^\ast$-algebra with unit $\1$: $\A$ is a {\it   quasi $^\ast$-algebra over $\Ao$}
if

\vspace{3mm}

{\bf [i]} the right and left multiplications of an element of $\A$
and an element of $\Ao$ are always defined and linear;


{\bf [ii]} $x_1 (x_2 a)= (x_1x_2 )a, (ax_1)x_2= a(x_1 x_2)$ and
$x_1(a
 x_2)= (x_1 a) x_2$, for each $x_1, x_2 \in \A_0$ and $a \in \A$;


{\bf [iii]} an involution * (which extends the involution of $\Ao$)
is defined in $\A$ with the property $(ab) ^\ast =b ^\ast a ^\ast$
whenever the multiplication is defined.



\vspace{2mm}

A quasi  $^\ast$ -algebra $(\A,\Ao)$ is {\it   locally convex} (or
{\it   topological}) if in $\A$ a { locally convex topology} $\tau$
is defined such that (a) the involution is continuous and the
multiplications are separately continuous; and (b) $\Ao$ is dense in
$\A[\tau]$.

Let $\{p_\alpha\}$ be a directed set of seminorms which defines
$\tau$. The existence of such a directed set can always be assumed.
We can further  also assume that $\A[\tau]$ is {\it complete}.

An explicit realization of a quasi *-algebra is constructed as follows: let $\Hil$ be a separable Hilbert space and $N$ an
unbounded, densely defined, self-adjoint operator. Let $D(N^k)$ be
the domain of the operator $N^k$, $k\in \N$, and $\D$ the domain of
all the powers of $N$:  $ \D \equiv D^\infty(N) = \cap_{k\geq 0}
D(N^k). $ This set is dense in $\Hil$. Let us now introduce
$\Lc^\dagger(\D)$, the *-algebra of all the {  closable operators}
defined on $\D$ which, together with their adjoints, map $\D$ into
itself. Here the adjoint of $X\in\Lc^\dagger(\D)$ is {
$X^\dagger=X^*_{\restr \D}$}.

In $\D$ the topology is defined by the following $N$-depending
seminorms: $\phi \in \D \rightarrow \|\phi\|_n\equiv \|N^n\phi\|,$
 $n\in \mathbb{N}_0$, while the topology $\tau_0$ in $\Lc^\dagger(\D)$ is introduced by the seminorms
{\normalsize$$ X\in \Lc^\dagger(\D) \rightarrow \|X\|^{f,k} \equiv
\max\left\{\|f(N)XN^k\|,\|N^kXf(N)\|\right\},\vspace{-2mm}$$} 

where
$k\in\mathbb{N}_0$ and   $f\in\C$, the set of all the positive,
bounded and continuous functions  on $\mathbb{R}_+$, which are
decreasing faster than any inverse power of $x$:
$\Lc^\dagger(\D)[\tau_0]$ is a {   complete *-algebra}.

Let further {  $\Lc(\D,\D')$} be the set of all continuous maps from
$\D$ into $\D'$, with their topologies (in $\D'$ this is the {
strong dual topology}), and let $\tau$ denotes the
topology defined by the seminorms $$ X\in \Lc(\D,\D') \rightarrow
\|X\|^{f} = \|f(N)Xf(N)\|,$$ $f\in\C$. Then $\Lc(\D,\D')[\tau]$ is a
{ complete vector space}.

In this case { $\Lc^\dagger(\D)\subset\Lc(\D,\D')$} and the pair
$$(\Lc(\D_,\D')[\tau],\Lc^\dagger(\D)[\tau_0])$$ is a {\it concrete realization} of a locally convex quasi
*-algebra.

\vspace{2mm}

We will discuss how to define a sort of {\em quantum dynamics} and a state over a {\em left} and a {\em right} version of $\LD$ which is a KMS state with respect to this dynamics. For that, as already stated, we first need to identify an ideal of $\LD$ which should play the same role that the Hilbert-Schmidt operators play in our previous construction. There was a lot of interest in ideals in algebras of unbounded operators, and many possible sets were proposed, \cite{timmideals} . In this section we will assume that $\D$ is a dense subset of the Hilbert space $\Hil$, and that an orthonormal basis of $\Hil$, $\F=\{\varphi_j,\,j\in I\}$, exists whose vectors all belong to $\D$: $\varphi_j\in\D$, $\forall j\in I$, where $I$ is a given set of indexes. We also assume that $\LD$ possesses the identity $\1$. Here we consider the following set
\be
\Nc=\left\{X\in\LD:\left|tr(AXB)\right|<\infty,\quad \forall A,B\in\LD\right\},
\label{31}\en
where $tr$ is the trace in $\Hil$. Of course $\Nc$ is a linear vector space. It is closed under the adjoint and it is an ideal for $\LD$: if $X\in\Nc$ then $AXB\in\Nc$, for all $A$ and $B$ in $\LD$. Hence, for each $X\in\Nc$, both $\left|tr(AXX^\dagger B)\right|$ and $\left|tr(AX^\dagger X B)\right|$ exist finite, for all $A$ and $B$ in $\LD$.

Let $\Phi,\Psi\in \D$. It is possible to check that the operator $X$ defined on $\Hil$ as $Xf=\left<\Psi,f\right>\Phi$ is an operator in $\Nc$, which using the Dirac's notation we will indicate as $X=|\Phi\left>\right<\Psi|$. Then the set $\Nc$ is rather rich. In particular it contains all the operators $X_{i,j}:=|\varphi_i\left>\right<\varphi_j|$, $i,j\in I$. We call $\G$ the set of all these operators. As for the Hilbert-Schmidt operators we can introduce on $\Nc$ a scalar product defined as
\be
\left<X,Y\right>_2=tr(X^\dagger Y),
\label{32}\en
for all $X, Y\in \Nc$. This map is well defined on $\Nc$ and, moreover, it is a scalar product. The set $\G$ is orthonormal with respect to this product. Indeed we find
\be
\left<X_{i,j},X_{k,l}\right>_2=\delta_{i,k}\delta_{j,l},
\label{33}\en
for all $i, j, k$, and $l$ in $I$. Moreover, $\G$ is complete in $\Nc$ with respect to this scalar product: indeed, the only element $Z\in\Nc$ which is orthogonal to all the $X_{i,j}$'s is $Z=0$.  We can also check that the only element of $\LD$ which is orthogonal to all the $X_{i,j}$'s is again the $0$ operator. In this sense, therefore, $\G$ is also complete in $\LD$.

The operators in $\Nc$ are very good ones. Indeed it is possible to check that, if $X\in\Nc$, then $X$ is bounded and $\|X\|\leq \|X\|_2$, where $\|.\|$ is the usual norm in $\Lc(\Hil)$. This means that $\Nc$ is a subset of the Hilbert space of the Hilbert-Schmidt operators, $\B_2(\Hil)$: $\Nc\subseteq \B_2(\Hil)$. However, it is still not clear wether
$\N$ is equal to $\B_2(\Hil)$ or not. This is part of the work in progress \cite{bit}.

\vspace{2mm}

Let us now define $\sigma_2=\overline{\Nc}^{\|.\|_2}$. This is an Hilbert space with scalar product $\left<.,.\right>_2$ which can, at most, coincide with $\B_2(\Hil)$. $\G$ is an orthonormal basis of $\sigma_2$.

As before we now introduce a map $\vee$ which, for all $A, B\in\LD$ and $X\in\Nc$, associates an element of $\Nc$:
\be
(A\vee B)(X):=AXB^\dagger
\label{34}\en

Of course, if $\Nc\subset\sigma_2$ then the problem of extending $\vee$ to $\sigma_2$ arises \cite{bit}.

This map, analogously to the one introduced in Section 1,  satisfies the following properties:
$$
(A\vee B)^\dagger = A^\dagger \vee B^\dagger,
  \quad
  (A_1\vee B_1) (A_2 \vee B_2) = (A_1A_2 ) \vee (B_1B_2)
$$
for all $A,B,A_1,B_1,A_2,B_2\in\LD$. Next, in analogy with what done before, we associate to $\LD$ two different algebras, $\LDl$ and $\LDr$ defined as follows:
$$
\LDl:=\left\{A_l:=A\vee\1,\quad A\in\LD\right\},\, \LDr:=\left\{A_l:=\1\vee A,\quad A\in\LD\right\}
$$
We also define  a map $J$ acting on $\G$ as follows: $J(X_{i,j})=X_{j,i}$, $i,j\in I$. Properties similar to those discussed in Section 1 can be recovered for $\LDl$, $\LDr$ and $J$, \cite{bit} .

Let now ${\lambda_i,\,i\in I}$ be a sequence of strictly positive numbers such that $\sum_{i\in I}\lambda_i=1$. We define a vector $\Phi\in\sigma_2$ as
\be
\Phi=\sum_{i\in I}\sqrt{\lambda_i} X_{i,i}.
\label{35}\en
We will assume here that, in particular, $\Phi\in \Nc$.

Using $\Phi$ we can now define two states over $\LDl$ and $\LDr$ as follows:
\be
\varphi(A_l):=\left<\Phi,A_l\Phi\right>_2=\left<\Phi,A\Phi\right>_2, \qquad A_l\in\LDl,
\label{36}\en and
\be
\tilde\varphi(A_r):=\overline{\left<\Phi,A_r\Phi\right>_2}=\left<\Phi,A\Phi\right>_2,  \qquad A_r\in\LDr.
\label{37}\en
We deduce that, if $A_r=\1\vee A$ and $A_l=A\vee\1$, $\tilde\varphi(A_r)=\varphi(A_l)$. Hence the two states are "very close" to each other. Therefore we will concentrate only on the properties of $\varphi$, since those of $\tilde\varphi$ are completely analogous and can be proved in the same way. First we observe that $\varphi$ can be written in a trace form by introducing the operator $\rho_\varphi:=\sum_{n\in I}\,\lambda_n\,X_{n,n}$. Indeed we find that
\be
\varphi(A_l)=tr(\rho_\varphi\, A), \qquad A_l\in\LDl.
\label{38}\en
This suggests to call $\varphi$, with a little abuse of notation, a {\em normal state} over $\LDl$.
We remark that $\rho_\varphi$ is bounded, $\|\rho_\varphi\|\leq \sum_{n\in I}\,\lambda_n=1$, and, more than this, it is  trace class, so that $\varphi$ is normalized: $\varphi(\1_l)=tr(\rho_\varphi)=\sum_{n\in I}\,\lambda_n=1$. We can also also check that $\rho_\varphi\in\sigma_2$, since $\left<\rho_\varphi,\rho_\varphi\right>_2=\sum_{n\in I}\,\lambda_n^2<\infty$. Moreover, $\rho_\varphi$ is self-adjoint and satisfies the eigenvalue equation $\rho_\varphi\varphi_j=\lambda_j\varphi_j$, $j\in I$.

The state $\varphi$ is positive and faithful on $\LDl$. Positivity is clear. As for faithfulness,  since for all $A_l\in\LDl$ $\varphi(A_l^\dagger A_l)=tr(\rho_\varphi A^\dagger A)=\sum_{n\in I}\,\lambda_n\,\|A\varphi_n\|^2$, then if $\varphi(A_l^\dagger A_l)=0$, the positivity of the $\lambda_n$'s implies that $A\varphi_n=0$ for all $n\in I$, so that $A=0$ and, as a consequence, $A_l=0$.

The vector $\Phi$ is cyclic for $\LDl$: let $X\in\Nc$ be orthogonal to $A_l\Phi$, for all $A_l\in\LDl$. Then we can show that $X=0$. Indeed we have, for all $A_l\in\LDl$
$$
0=\left<X,A_l\Phi\right>_2=\sum_{n\in I}\lambda_n\,\left<\varphi_n,X^\dagger A\varphi_j\right>.
$$
Hence, in particular, $\left<X,A_l\Phi\right>_2=0$ if $A_l=A\vee\1$ with $A=X_{k,l}$, for $k,l\in I$. But, since $\lambda_n>0$ for all $n\in I$, $\left<X\varphi_l,\varphi_k \right>=0$ for all $k,l\in I$. Then $X=0$.

The vector $\Phi$ is also separating for $\LDl$. Suppose in fact that $A_l\Phi=0$. Then we can prove that $A_l=0$. This follows from the fact that $0=\left<A_l\Phi,A_l\Phi\right>_2=\sum_{n\in I}\,\lambda_n\,\|A\varphi_n\|^2$, which implies again that $A=0$ and $A_l=0$.

\vspace{2mm}

We can now define an operator $P_{i,j}$ mapping $\Nc$ into $\Nc$ in the following way:
\be
P_{i,j}:=X_{i,i}\vee X_{j,j}
\label{39}\en
It is possible to check that $P_{i,j}^\dagger=P_{i,j}$ and that $P_{i,j}P_{k,l}=\delta_{i,k}\delta_{j,l}P_{i,j}$. Let us further introduce an operator $H_\varphi$ acting on $\F$ as follows:
\be
H_\varphi\,\varphi_j=-\frac{1}{\beta}\,\log(\lambda_j)\,\varphi_j,
\label{310}\en
$j\in I$. Here $\beta$ is a positive constant which, in the physical literature, is usually called {\em the inverse temperature}. Then $H_\varphi$ can be written as $H_\varphi=-\frac{1}{\beta}\,\sum_{j\in I}\log(\lambda_j)\,X_{j,j}$, where, of course, the sum converges on the dense domain of all the finite linear combinations of the vectors of $\F$.

Via the spectral theorem we deduce that $\rho_\varphi$ and $H_\varphi$ are related to each other. Indeed we have $\rho_\varphi=e^{-\beta H_\varphi}$. Extending the map $\vee$ we can also define the following operators:
\be
H_\varphi^l=H_\varphi\vee\1; \quad H_\varphi^r=\1\vee H_\varphi; \quad h_\varphi=H_\varphi^l-H_\varphi^r
\label{311}\en
which can be written as
\be\begin{array}{ll}
H_\varphi^l=-\frac{1}{\beta}\,\sum_{i,j\in I}\log(\lambda_i)\,P_{i,j}; \quad H_\varphi^r=-\frac{1}{\beta}\,\sum_{i,j\in I}\log(\lambda_j)\,P_{i,j}; \\
h_\varphi=-\frac{1}{\beta}\,\sum_{i,j\in I}\log\left(\frac{\lambda_i}{\lambda_j}\right)\,P_{i,j}.\end{array}
\label{312}\en
In particular $h_\varphi$ can be used to define the operator $\Delta_\varphi$ as follows:
\be
\Delta_\varphi=e^{-\beta h_\varphi}=\sum_{i,j\in I}\,\frac{\lambda_i}{\lambda_j}\,P_{i,j},
\label{313}\en
and $\Delta_\varphi$ can be used to define a {\em Schr\"odinger dynamics}:
\be
e^{i\,t\, h_\varphi}=\left(\Delta_\varphi\right)^{-it/\beta}=\sum_{i,j\in I}\,\left(\frac{\lambda_i}{\lambda_j}\right)^{-it/\beta}\,P_{i,j}.
\label{314}\en
It is possible to check that $e^{ith_\varphi}=e^{itH_\varphi}\vee e^{itH_\varphi}$ and that $e^{ith_\varphi}(\Phi)=\Phi$, which is therefore invariant under our {\em time evolution}. Using $e^{ith_\varphi}$ we can also define an Heisemberg dynamics on $\LDl$:
\be
\alpha_\varphi^t(A_l)=e^{i\,t\, h_\varphi}\,A_l\,e^{-i\,t\, h_\varphi}=\left(e^{i\,t\, H_\varphi}\,A\,e^{-i\,t\, H_\varphi}\right)\vee\1
\label{315}\en
If $e^{i\,t\, H_\varphi}\in\LD$, $\alpha_\varphi^t(A_l)$ belongs to $\LDl$ for all $A_l\in\LDl$. The invariance of $\Phi$ under the action of $e^{ith_\varphi}$ implies that $\varphi(\alpha_\varphi^t(A_l))=\varphi(A_l)$ for all $A_l\in\LDl$. Moreover, we can also check that $\varphi$ is a KMS state with respect to the time evolution $\alpha_\varphi^t$.

\vspace{3mm}

Of course, now the next step is to apply this procedure to some concrete example, like the one we have considered in Section 2. Particularly interesting for us will be consider some concrete application of the procedure to $QM_\infty$. This is  work in progress, \cite{bit} .

\section*{Acknowledgements}

I want to express my gratitude to the local organizers of the conference, and in particular to Prof. Rolando Rebolledo, for their patience.

\end{document}